\documentclass[apjl]{emulateapj}

\usepackage{amsmath}
\usepackage{txfonts}
\usepackage{natbib}
\usepackage{graphicx}
\usepackage{color} 

\begin{document}

\title{Progenitor-explosion connection and remnant birth masses for 
neutrino-driven supernovae of iron-core progenitors}

\shorttitle{Progenitor-remnant connection for neutrino-driven supernovae}

\shortauthors{Ugliano et al.}

\author{Marcella Ugliano \altaffilmark{1},
  Hans-Thomas Janka \altaffilmark{1},
  Andreas Marek \altaffilmark{1}, and
  Almudena Arcones \altaffilmark{2,3}
  }

\altaffiltext{1}{Max-Planck-Institut f\"ur Astrophysik,
       Karl-Schwarzschild-Str. 1, D-85748 Garching, Germany}

\altaffiltext{2}{Institut f\"ur Kernphysik,
Technische Universit\"at Darmstadt,
Schlossgartenstr.~2,
D-64289 Darmstadt,
Germany}

\altaffiltext{3}{GSI  Helmholtzzentrum
f\"ur Schwerionenforschung GmbH,
Planckstr.~1,
D-64291 Darmstadt,
Germany}

\begin{abstract}
We perform hydrodynamic supernova simulations in spherical symmetry
for over 100 single stars of solar metallicity to explore the
progenitor-explosion and progenitor-remnant connections established
by the neutrino-driven mechanism. We use an approximative treatment 
of neutrino transport and replace the high-density interior of the 
neutron star (NS) by an inner boundary condition based on an 
analytic proto-NS core-cooling model, whose
free parameters are chosen such that explosion energy, nickel
production, and energy release by the compact remnant of progenitors
around 20\,$M_\odot$ are compatible with Supernova~1987A.
Thus we are able to simulate the accretion phase, initiation of the 
explosion, subsequent neutrino-driven wind phase for 15--20\,s, and the
further evolution of the blast wave for hours to days until fallback
is completed. Our results challenge long-standing paradigms. We find that
remnant mass, launch time, and properties of the explosion depend strongly
on the stellar structure and exhibit large variability even in narrow
intervals of the progenitors' zero-age-main-sequence mass.
While all progenitors with masses below $\sim$15\,$M_\odot$ yield
NSs, black hole (BH) as well as NS formation is possible for more massive
stars, where partial loss of the hydrogen envelope leads to weak
reverse shocks and weak fallback. Our NS baryonic masses of 
$\sim$1.2--2.0\,$M_\odot$ and BH masses $>$6\,$M_\odot$
are compatible with a possible lack of low-mass BHs in the empirical
distribution. Neutrino heating accounts for SN energies
between some $10^{50}$\,erg and $\sim$$2\times 10^{51}$\,erg, but
can hardly explain more energetic explosions and nickel masses
higher than 0.1--0.2\,$M_\odot$. These seem to require 
an alternative SN mechanism.
\end{abstract}


\section{Introduction}

One of the main goals of the long-standing quest for a better understanding
of the supernova (SN) explosion mechanism is the establishment of
a theoretical connection between the properties of progenitor stars 
and those of SNe and their remnants. Understanding the progenitor-remnant
systematics would not only mean a better definition of SNe
as the end point of massive-star evolution, it could also
serve as input for continuative theoretical studies like binary
population synthesis calculations. Moreover, it would provide indispensable 
information and the theoretical basis for deducing constraints from
observations on the physical processes that trigger and power the
onset of the explosion at the center of the SN. In particular, the
explosion mechanism plays a crucial role for determining the 
energetic and dynamical properties of the SN blast wave. It thus
controls the conditions for explosive nucleosynthesis and is of
central importance for answering long-standing astrophysical questions:
Which progenitors give birth to neutron stars (NSs), which ones to 
black holes (BHs)? How do SN properties depend on the progenitor
metallicity? Are the observed masses of compact remnants compatible
with modeling results? Do observed SNe provide evidence for more than
one explosion mechanism? In particular, considering neutrino heating
as the most widely favored process to trigger the 
explosions \citep{Bethe1985}, the 
specific question arises, which range of SN energies and nickel
masses can be explained by the neutrino-driven mechanism.

A popular way to model the blast wave initiation for studying 
SN nucleosynthesis and determining compact remnant masses are
piston-driven explosions \citep[e.g.,][and references
  therein]{Woosley1995,Zhang2008}, in which a piston is
placed at a chosen position in the pre- or post-collapse stellar 
core and moved with a prescribed, time-dependent velocity to mimic 
the infall of the selected mass shell and the explosive expansion
of the newly formed SN shock. Besides the initial mass cut ---fallback
of nonescaping material may shift the final mass cut--- also the
energy deposited by the piston is prescribed as a free parameter. 
Alternatively, internal energy ``bombs'' have been used, in which
a defined amount of thermal energy is released by an artificial
increase of the temperature in a certain volume \citep{Aufderheide1991}. 

Investigating spherically symmetric stellar core collapse by
hydrodynamic simulations with simplified neutrino treatment,
\citet{OConnor2011} identified a single parameter, the
``progenitor bounce compactness'' $\xi_{2.5}$ of the innermost
2.5\,$M_\odot$ of a star, whose value above a certain threshold 
they considered to be indicative for an unlikely explosion
and thus BH formation. Progenitors with $\xi_{2.5} \gtrsim 0.45$
were found to reach the BH formation limit typically within
$\lesssim$0.8\,s after bounce because of their tremendous mass
accretion rates. Such stars also turned out to require a particularly 
high time-averaged 
neutrino-heating efficiency to develop an explosion against the
huge ram pressure of the infalling core material. Nevertheless, the
exact value of $\xi_{2.5}$ where the bifurcation of the behavior
occurs, appears somewhat arbitrarily chosen and not supported by
strong theoretical arguments. In particular, the BH formation limit
could already take place at a lower value of $\xi_{2.5}$.

\citet{Belczynski2011} and \citet{Fryer2012} made theoretical
predictions of the NS/BH mass distribution by 
applying a simple analytic concept for estimating the explosion time,
explosion energy, and fallback mass on the basis of
structural properties of the progenitor stars \citep{Fryer2006}.
\citet{Fryer2001} derived theoretical BH mass distributions by employing the explosion 
energy as function of the progenitor mass from a fit to results
of a small set of four 2D models of neutrino-driven explosions 
\citep{Fryer1999}. They considered different fallback masses in
dependence on the (parametized) fraction of the energy released
by the SN mechanism that is used to unbind the outer layers of
the progenitor stars.

In this work we adopt a different approach. We perform
hydrodynamical simulations in spherical symmetry for a large set
of 101 progenitor models, following core collapse and bounce, postbounce 
accretion, the possible onset of an explosion and subsequent neutrino-driven
wind phase of the cooling proto-neutron star (PNS), and the development of
the SN blast wave over hours to days until the fallback mass is determined. 
In order to power the explosions, we refer to the neutrino-driven mechanism,
whose viability seems to be supported by recent 2D simulations with detailed
energy-dependent neutrino transport \citep{Marek2009} and general
relativity \citep{Muller2012a,Muller2012b},
although an ultimate robustness of the model predictions may require
3D simulations and may hinge on remaining uncertainties of the
high-density neutrino opacities and equation of state (EoS).
Since in our spherical simulations explosions by neutrino heating must be 
triggered artificially, we excise the high-density core of the nascent NS
and replace it by a time-dependent, analytic cooling model. This core model
yields the time evolution of the neutrino emission from the high-density
interior of the PNS. Its free parameters are adjusted such that
observational characteristics of SN~1987A, i.e., the explosion energy, 
$E_{\mathrm{87A}} \sim 1.3\times 10^{51}$\,erg, 
and ejected $^{56}$Ni mass, $M_{\mathrm{Ni,87A}} \sim 0.07\,M_\odot$,
are reproduced for progenitor stars around 20\,$M_\odot$
\citep[e.g.,][]{Tanaka2009,Utrobin2011}. This implies the
assumption that SN~1987A was an ordinary case of a neutrino-driven 
explosion in which, e.g., exceptionally rapid core rotation did not
play a crucial role for the collapse dynamics. The NS core model with 
the same parameter values is then also applied to all other collapsing
stars. 

Provided SN~1987A was such a normal case and neutrino-energy deposition
was responsible for its explosion, our approach thus allows us to explore the 
response of other progenitor stars to the input of neutrino energy released
from their forming compact remnants, and in the case of successful explosions
to determine the blast wave and remnant properties of the stars.
In the following we will give more information on our modeling strategy
and tools, present our results, and discuss the implications.

\section{Numerical setup and input}
\label{sec:numerics}

We employ the explicit, finite-volume, Eulerian multi-fluid
code {\sc Prometheus} \citep{Fryxell1989} to solve the
1D hydrodynamics equations on a spherical grid. We use typically
1000 geometrically spaced radial zones with the outer boundary
at a radius $R_\mathrm{ob}$. The latter is placed at a large 
distance (initially at 150,000\,km) that is not directly 
affected by the simulated stellar collapse and SN explosion. In
particular it is sufficiently large to prevent the SN shock from 
leaving the computational domain during the simulated time.
The advection of nuclear species is treated by the Consistent
Multi-fluid Advection (CMA) scheme of \citet{Plewa1999}, and self-gravity
of the stellar plasma includes corrections of the Newtonian
gravitational potential for general relativistic effects as
described in detail in \citet{Scheck2006} and \citet{Arcones2007}.

The inner core of the PNS with a mass of $M_\mathrm{c}=1.1\,M_\odot$
and densities well above those of the neutrinospheric layer is 
excised and replaced by a point mass at the coordinate origin.
The shrinking of
the PNS is mimicked by a retraction of the closed inner boundary at
$R_\mathrm{ib}$ together with the radial grid, applying the function
defined in \citet{Arcones2007}. For the simulations discussed in
this paper we use the values $R_\mathrm{ib}^\mathrm{i} = 
R(1.1\,M_\odot\,\,\mathrm{at}\,\,t$$\sim$10\,ms p.b.)\,$\sim 60$\,km,
$R_\mathrm{ib}^\mathrm{f} = 20$\,km, and $t_\mathrm{ib} = 0.4$\,s
for the initial and final boundary radii and the exponential
contraction timescale, respectively. The chosen values yield a 
boundary contraction similar to the ``standard case'' of 
\citet{Scheck2006}. Hydrostatic equilibrium is assumed at the
boundary radius $R_\mathrm{ib}$ between excised core and accretion 
envelope of the nascent NS.

The cooling of the PNS interior and corresponding boundary luminosities
are described by the neutrino core-cooling model presented in 
Sect.~\ref{sec:coremodel}. 
The temperatures of the spectra of inflowing neutrinos (assumed to
have Fermi-Dirac shape) are set equal to the gas temperature
$T_\mathrm{i}$ in the innermost grid zone, 
$T_\nu(R_\mathrm{ib}) = T_\mathrm{i}$,
consistent with the fact that neutrinos in highly optically thick
regions are close to local thermal equilibrium.
The spectral degeneracy parameters are chosen to be
$\eta_{\nu_e} = \eta_{\bar\nu_e} = \eta_{\bar\nu_x} = 0$
for electron neutrinos $\nu_e$, electron antineutrinos $\bar\nu_e$,
and heavy-lepton neutrinos $\nu_x$ (= $\nu_\mu$, $\bar\nu_\mu$,
$\nu_\tau$, $\bar\nu_\tau$), respectively. (Using the values for
local chemical equilibrium for $\eta_{\nu_e}$ and 
$\eta_{\bar\nu_e}$ does not cause any significant differences.)
Neutrino transport and neutrino-matter interactions in the
computational domain are
approximated as in \citet{Scheck2006} by radial integration of the
one-dimensional (spherical), grey transport equations for neutrino
number and energy, assuming the neutrino spectra to have Fermi-Dirac 
shape with a calculated local temperature $T_\nu$.

The high-density plasma is described
by the tabulated microphysical equation of state (EoS) of
\citet{Janka1996} including arbitarily degenerate and
arbitarily relativistic electrons and positrons, photons, and
four predefined nuclear species (neutrons, protons, alpha 
particles, and a representative Fe-group
nucleus) in nuclear statistical equilibrium (NSE).
At temperatures and densities below NSE the thermodynamics are
treated by the Helmholtz-EoS of \citet{Timmes2000}.

The NS cooling and its neutrino emission are followed for
8--25\,s until the shock has crossed $10^{10}$\,cm.
At that time the power of the neutrino-driven
wind, which pushes the dense ejecta shell enclosed by the expanding
SN shock and the wind-termination shock, has decayed to a dynamically
insignificant level.
The models are then mapped to a grid extending to larger radii, and
the inner boundary is moved from below the neutrinosphere to $10^9$\,cm
in order to ease the time step constraint. The gravity potential of
the central object is thereby corrected for the gas mass in the larger
excised region, and free-outflow conditions are applied at the new
inner boundary. When the SN blast is followed beyond shock breakout for
several days, the star is embedded in a wind-like medium with a density
$\rho\propto r^{-2}$ and a constant temperature of $T = 1100$\,K, and
the grid is shifted outward once again to cover radii from
$R_\mathrm{ib} = 10^{10}$\,cm to $R_\mathrm{ob}$ up to $10^{15}$\,cm.

During the long-time simulations of the SN dynamics we follow
approximately the explosive nucleosynthesis 
by solving a small alpha-reactions network, similar to
the network described in \citet{Kifonidis2003}. It consists
of the 13 $\alpha$-nuclei: $^4$He, $^{12}$C, $^{16}$O,
$^{20}$Ne, $^{24}$Mg, $^{28}$Si, $^{32}$S, $^{36}$Ar, 
$^{40}$Ca, $^{44}$Ti, $^{48}$Cr, $^{52}$Fe, $^{56}$Ni, and
an additional tracer nucleus, which is 
produced via the reaction rate for
$^{52}$Fe($\alpha$,$\gamma$)$^{56}$Ni within grid cells whose
electron fraction $Y_e$ is below 0.49. This allows us to keep
track of element formation in regions with neutron excess. The network
is solved in grid cells whose temperature is between
$10^8$\,K and $7\times 10^9$\,K. 
We assume that at $T > 7\times 10^9$\,K all nuclei are
photo-disintegrated to $\alpha$-particles. 
Such a composition is 
consistent with the NSE yields that are produced by our
network solver in the high-temperature limit, because the burning
network contains $\alpha$-particles as the only representatives
of light nuclear species.
A feedback from the network composition to the EoS
and thus to the hydrodynamic evolution is neglected.

\subsection{Proto-neutron star core model}
\label{sec:coremodel}

The cooling of the PNS core is described by an analytic model that
couples the excised core region to the surrounding accretion layer,
whose evolution is followed on the computational grid. 
We assume the dense PNS core
with mass $M_\mathrm{c}$ and radius $R_\mathrm{c}$ to be approximately
homogeneous and its EoS to be an ideal $\Gamma$ law, $P = (\Gamma-1)e$
(with $P$ being the pressure, $e$ the internal energy density).
Combining the total core energy, 
$E_\mathrm{c} = E_\mathrm{g} + E_\mathrm{i}$, and the virial theorem,
$E_\mathrm{g} + 3(\Gamma-1)E_\mathrm{i} + S = 0$, we can replace 
the integrated internal energy, $E_\mathrm{i}$, and express
$E_\mathrm{c}$ in terms of the Newtonian gravitational energy,
$E_\mathrm{g} = -\frac{2}{5} GM_\mathrm{c}^2/R_\mathrm{c}$, and the
surface term, $S = -4\pi R_\mathrm{c}^3 P_\mathrm{s}$, 
for pressure $P_\mathrm{s}$ at $R_\mathrm{c}$:
\begin{equation}
E_\mathrm{c} = \frac{3\Gamma-4}{3(\Gamma-1)}\,E_\mathrm{g}
 - \frac{S}{3(\Gamma-1)} \,.
\label{eq:energyfunc}
\end{equation}
The energetic evolution of the core is given by its loss of 
neutrino energy and compression work done on its surface as
\begin{equation}
\dot E_\mathrm{c} \equiv \frac{\mathrm{d}E_\mathrm{c}}{\mathrm{d}t} =
- L_{\nu,\mathrm{c}} - 
4\pi P_\mathrm{s} R_\mathrm{c}^2 \dot R_\mathrm{c} \, ,
\label{eq:energychange}
\end{equation}
where $\dot E_\mathrm{c}$ can be computed as time derivative of
Eq.~(\ref{eq:energyfunc}), $L_{\nu,\mathrm{c}}$ is the total neutrino 
luminosity, and the second term results from the time derivative of
the core volume. Instead of setting $P_\mathrm{s}$ equal to the 
boundary pressure on the hydro grid, we prefer to link it to overall
properties of the accretion layer. This prescription is intended to
capture the nature of the core-mantle coupling but not to constrain
the freedom to tune the parameters of the simple PNS core model. 
We therefore consider hydrostatic equilibrium in terms of the 
mass coordinate $m(r)$,
$\mathrm{d}P/\mathrm{d}m = - GM/(4\pi r^4)$, and
linearize both sides to obtain
\begin{equation}
P_\mathrm{s} = \zeta\,
\frac{G M_\mathrm{c} m_\mathrm{acc}}{4\pi R_\mathrm{c}^4} \, ,
\label{eq:ps}
\end{equation}
where $m_\mathrm{acc}$ is the mass of the accretion layer
that surrounds the PNS core, $\zeta > 0$ a numerical factor of order
unity, and we assumed $P_0 \ll P_\mathrm{s}$ for the pressure $P_0$
outside of the accretion layer. Moreover, in performing the time 
derivative of Eq.~(\ref{eq:energyfunc}) we assumed $M_\mathrm{c}$
and $\Gamma$ to be constant. Combining 
Eqs.~(\ref{eq:energyfunc})--(\ref{eq:ps}) we arrive at
\begin{equation}
L_{\nu,\mathrm{c}} = \frac{3\Gamma -4}{3(\Gamma-1)}\,
(E_\mathrm{g} + S)\,\frac{\dot R_\mathrm{c}}{R_\mathrm{c}} -
\frac{\zeta}{3(\Gamma-1)}\,\frac{\delta E_\mathrm{acc}}{\delta t}
\label{eq:nulum}
\end{equation}
with $S = - \zeta\,GM_\mathrm{c}m_\mathrm{acc}/R_\mathrm{c}$ and
$\delta E_\mathrm{acc}/\delta t \equiv 
GM_\mathrm{c}\dot m_\mathrm{acc}/R_\mathrm{c}$. 
While the first term on the rhs of
Eq.~(\ref{eq:nulum}) describes the luminosity increase due to
the deepening of the gravity potential and surface work in the
case of PNS compression, the second term accounts for the higher
core pressure (and internal energy) needed when the accretion layer
grows in mass.

Equation~(\ref{eq:nulum}) is used to prescribe the boundary 
luminosities, 
$L_{\nu_i,\mathrm{c}} \equiv \ell_{\nu_i} L_{\nu,\mathrm{c}}$, 
of neutrinos of all kinds with $\ell_{\nu_e} = 0.20$, 
$\ell_{\bar\nu_e} = 0.15$, $\ell_{\nu_x} = 0.1625$. (This choice
corresponds to a certain loss of lepton number from the 
PNS core, see \citet{Scheck2006}, and ensures a reasonable
evolution of $Y_e$ in the PNS mantle and surface layers, but does 
not have much relevance for the dynamical evolution.) 

In our simulations $m_\mathrm{acc}$ is taken to
be the mass between the inner grid boundary and a density of 
$\rho_0 = 10^{10}\,$g\,cm$^{-3}$ at radius $r_0$, where we 
define $\dot m_\mathrm{acc} = -4\pi r_0^2 v_0 \rho_0$ (accretion 
means a velocity $v_0 < 0$ and $\dot m_\mathrm{acc} > 0$). 
The core radius is assumed to 
contract according to $R_\mathrm{c}(t) = R_\mathrm{c,f} + 
(R_\mathrm{c,i}-R_\mathrm{c,f})/(1 + t)^n$ with
$R_\mathrm{c,i} = R_\mathrm{ib}^\mathrm{i}$ and 
$R_\mathrm{c,f}$ being the initial and final radius, respectively,
and $t$ is measured in seconds. With $\Gamma = 3$, $n = 3$,
and $\zeta = 0.6$ a choice of $R_\mathrm{c,f} = 6\,$km allows us 
to reproduce $E_\mathrm{87A}$ and $M_\mathrm{Ni,87A}$ of SN~1987A 
for progenitors in the 20\,$M_\odot$ range. For the simulations
discussed below the 19.8\,$M_\odot$ progenitor serves for the
calibration of the PNS core model, but the overall results are 
similar when neighboring stars or a SN~1987A blue supergiant
progenitor \citep{Woosley1988} are used for the calibration.

\begin{figure}
\plotone{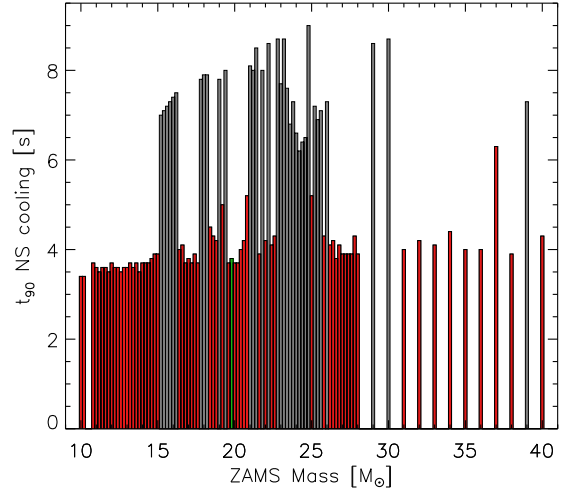}
\caption{Timescale of 90\% of the neutrino-energy loss of the
forming compact remnant as function of the progenitor ZAMS mass.
Red histogram bars indicate successful explosions, grey ones
correspond to cases where BHs form without a SN
explosion, and the green bar marks the 19.8\,$M_\odot$ progenitor
used for the calibration with SN~1987A observations (see text).
The BH formation cases correspond to ``cooling times''
in excess of 6\,s, because the compact object in our simulations
remains radiating neutrinos even
when its mass nominally exceeds the BH formation limit. This
implies that our modeling does not invoke any assumption about the
equation-of-state dependent mass limit for BH formation.}
\label{fig:pnscoolingtime}
\end{figure}

The chosen parameter values lead to typical PNS neutrino-cooling 
times ($t_{90}$ for 90\% of the total neutrino-energy release) 
of 3.5--5.5\,s (Fig.~\ref{fig:pnscoolingtime}).
This is shorter than the $\sim$10\,s of emission inferred
from the SN~1987A neutrino events of Kamiokande~II. However, this
detector reported a 7\,s gap after 8 events in the first two
seconds, and the last 3 events were very close to the detection
threshold \citep{Hirata1987}. It is interesting to note
that the neutrino signal in all three experiments (Kamiokande~II,
Irvine-Michigan-Brookhaven, and Baksan) is 
compatible with a PNS cooling period (exponential
cooling timescale) of only 4--5\,s \citep{Raffelt1996,Loredo2002,Pagliaroli2009}.
We emphasize that our 1D simulations, which ignore postshock
convection and yield standard values for the mean energies of 
radiated $\nu_e$ and $\bar\nu_e$
($\left\langle\epsilon\right\rangle \approx$\,10--17\,MeV),
naturally require overestimated postbounce neutrino fluxes 
to trigger explosions. Since the neutrino emission of our PNS
model is roughly compatible with the SN~1987A data, we expect
that our calibration by SN~1987A explosion properties reflects
trends of neutrino-driven explosions whose validity for the
investigated progenitors in relative comparison holds beyond
the considered 1D setup.
We also stress that our description of the core behavior accounts
for the presence of an accretion layer surrounding the (excised)
high-density core. The growth of this layer depends on the 
progenitor-specific stellar structure and corresponding mass-infall 
rate. Its accretion luminosity, which adds to the analytically
modeled core-neutrino luminosity, is also included in our simulations
by means of a simple, grey transport approximation.

The progenitor structure therefore influences
the postbounce evolution in different ways. The density profile of 
the outer Fe-core, Si-, and O-layers does not only
determine the mass accretion rate and thus the accretion luminosity;
it also governs the shock stagnation radius because high/low 
accretion rates damp/favor shock expansion. Moreover, according to
Eq.~(\ref{eq:nulum}) our PNS-core model connects the neutrino
emission of the PNS core to the growth of the accretion mantle
of the PNS. Core and accretion luminosity conspire in reviving and
powering the shock wave, for which reason the neutrino emission
properties of the PNS (core and accretion contributions) are crucial
for the explosion. It is therefore clear that the calibration of the
PNS-core parameters by the explosion properties of a 
chosen progenitor model cannot be independent of the structure of 
this progenitor star. The fact that basic aspects of our results
exhibit some robustness against variations of the 
calibration model is assuring. However, this does not imply
that the predicted explosion properties could not change if the
core structure of the SN~1987A progenitor was considerably
different from the $\sim$20$\,M_\odot$ solar metallicity 
stars and the blue supergiant model used in our calibration
tests. Metallicity
effects, interaction in a binary scenario, or rotation and
convective boundary effects could have affected the mass-loss
history and thus could have altered the growth of the stellar
core compared to the progenitors models considered by us for
calibration.

\begin{figure}
\plotone{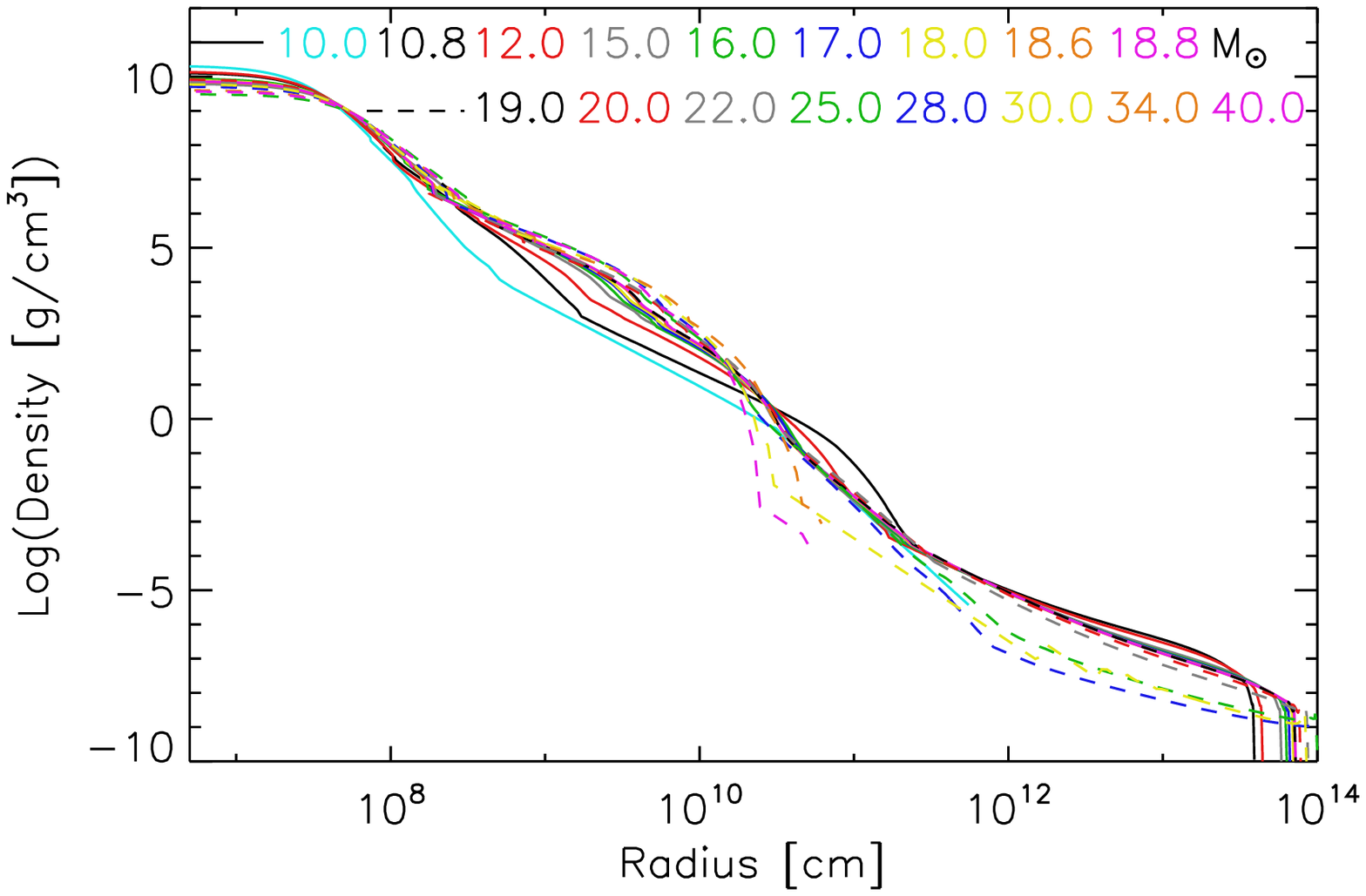}\\\vspace{3pt}
\plotone{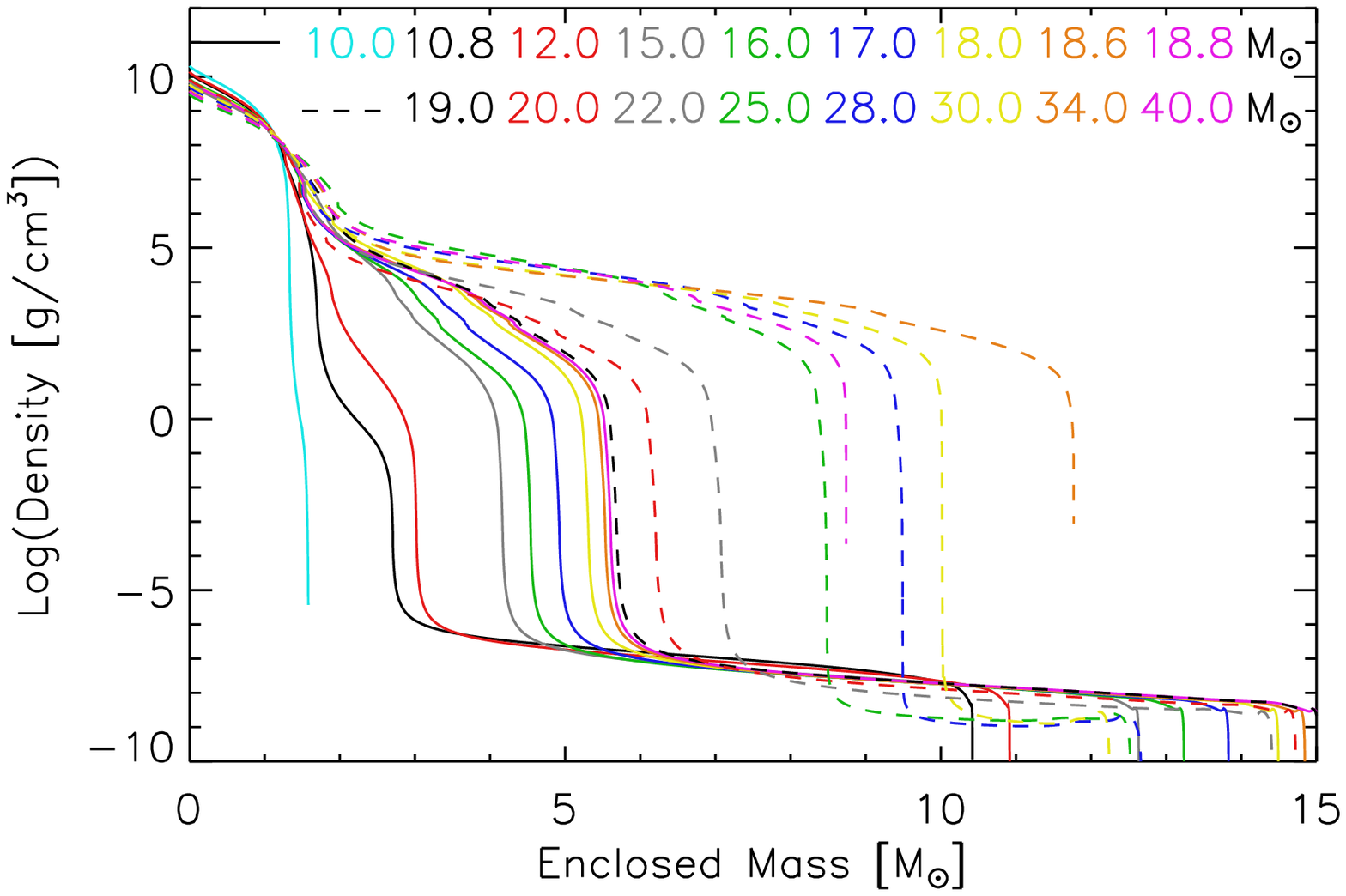}
%
\caption{Density profiles vs.\ radius ({\em top}) and vs.\ enclosed mass
({\em bottom}) at the onset of core collapse for a selection of models
from the considered set of solar-metallicity progenitors with iron
cores. Solid lines correspond to ZAMS masses less than 19\,$M_\odot$,
dashed lines to higher values. Note that the stellar shell structure
and also the high-density core
($\rho \gtrsim 10^5$\,g\,cm$^{-3}$) exhibit considerable variations
with the progenitor mass (see also Fig.~\ref{fig:progenitorprops}).}
\label{fig:progenitorprofs}
\end{figure}

\begin{figure}
\plotone{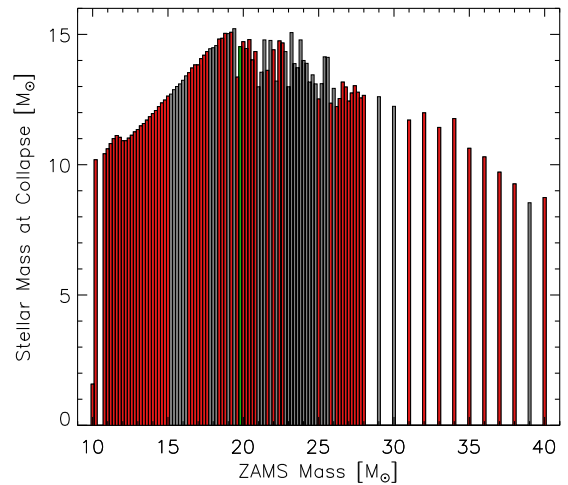}
\caption{Stellar masses at the onset of core collapse as function of 
ZAMS mass. Mass loss is particularly strong beyond $\sim$20\,$M_\odot$.
The red histogram bars indicate successful explosions, the grey shaded
ones correspond to cases where BHs form without
a SN explosion, and the green bar marks the 19.8\,$M_\odot$ progenitor
used for the calibration with SN~1987A observations (see text).}
\label{fig:massatcollapse}
\end{figure}

\begin{figure*}
\plotone{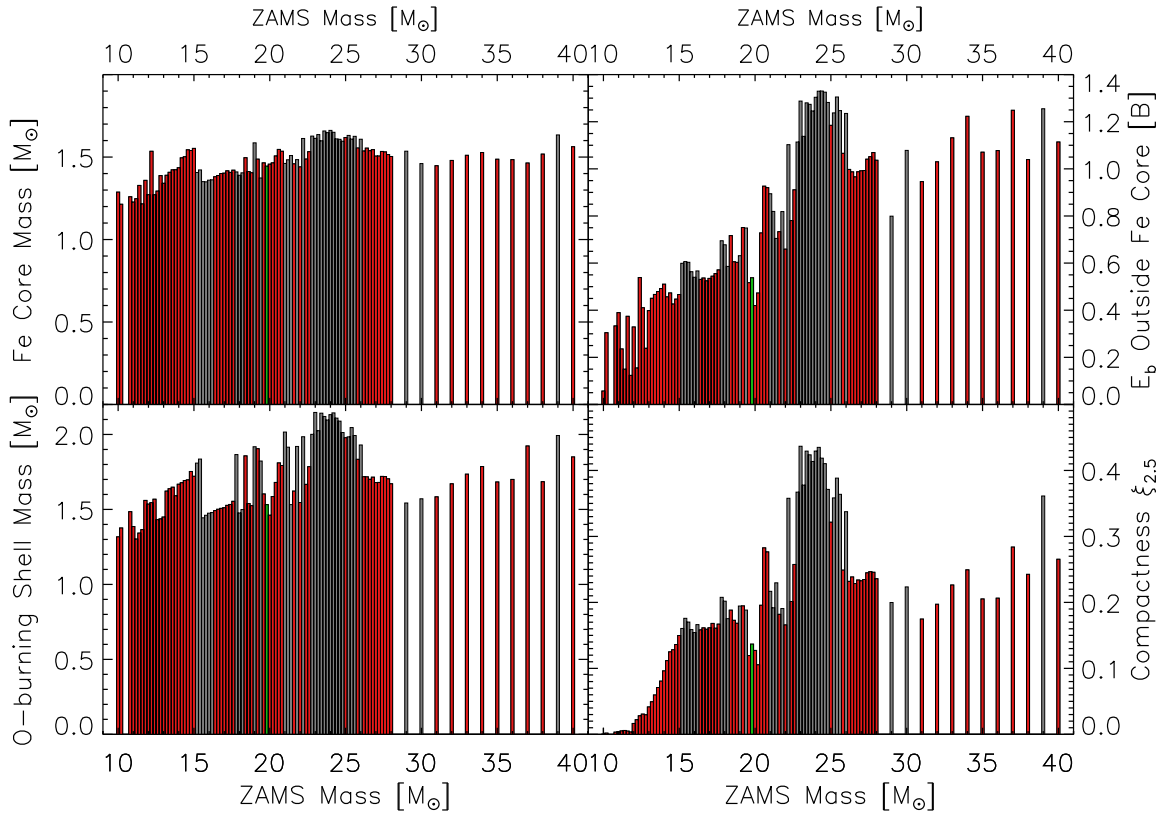}
%
\caption{Quantities characterizing the core structure of the progenitors
at the onset of collapse vs.\ ZAMS mass: deleptonized (``iron'') 
core mass ({\em top left}),
binding energy of the matter outside of the iron core in units of
1\,B = 1\,bethe = $10^{51}$\,erg ({\em top right})
at the onset of collapse, enclosed mass at the bottom of the oxygen-burning
shell ({\em bottom
left}), and compactness parameter $\xi_{2.5}$ of the innermost 
2.5\,$M_\odot$ as defined in Eq.~(\ref{eq:compactness})
({\em bottom right}). Red, grey, 
and green histogram bars have the same meaning as in
Fig.~\ref{fig:massatcollapse}.}
\label{fig:progenitorprops}
\end{figure*}

\subsection{Progenitor stars}
\label{sec:progenitors}

Density profiles versus radius and mass for a subset of 
the investigated 101 solar-metallicity progenitors with iron
cores \citep{Woosley2002} are displayed in 
Fig.~\ref{fig:progenitorprofs}. The models are given in 0.2\,$M_\odot$
steps between 10.8\,$M_\odot$ and 28\,$M_\odot$ and further up to 
40\,$M_\odot$ in 1.0\,$M_\odot$ steps. We also include two stars
with 10.0\,$M_\odot$ and 10.2\,$M_\odot$. 
While the density profiles are labeled 
with the zero-age-main-sequence (ZAMS) masses, the corresponding 
pre-collapse masses are visible in Fig.~\ref{fig:massatcollapse}.
The images show that mass loss is moderate up to 19.6\,$M_\odot$ 
(except for the 10.0\,$M_\odot$ case) but grows considerably 
beyond that value. 

The nonmonotonic star-to-star scatter is a consequence of the
turn-on and -off of convective shells during the late evolution
stages. The scatter can be seen in Fig.~\ref{fig:progenitorprops} 
for different structural parameters as functions of the ZAMS mass:
The iron-core mass, $M_\mathrm{Fe}$, is identified with the 
deleptonized core, i.e., the volume
where the electron faction $Y_e < 0.497$;
the binding energy of the overlying shells, $E_\mathrm{b}$,
is evaluated at the onset of core collapse, and the
enclosed mass at the inner edge of the oxygen-burning
shell, $M_{\mathrm{O}_<}$, is defined by the point where the
dimensionless entropy per nucleon reaches a value of 4. Finally,
the compactness parameter $\xi_{2.5}$ is defined as in Eq.~(10)
of \citet{OConnor2011} by the ratio of mass $M = 2.5\,M_\odot$
and radius $R(M)$ that encloses this mass:
\begin{equation}
\xi_{2.5} \equiv \frac{M/M_\odot}{R(M)/1000\,\mathrm{km}}\, ,
\label{eq:compactness}
\end{equation}
where we consider $M$ as the baryonic mass. For our solar-metallicity
progenitors it makes no noticeable difference whether we compute 
$\xi_{2.5}$ at the onset of collapse or at the time of bounce
\citep[as suggested by][]{OConnor2011}, in contrast to the 
situation for low-metallicity stars, which experience less mass
loss during their evolution and which develop more compact cores.

The grey histogram bars mark the cases where we found BH formation
without a SN explosion. No clear correlation can be observed with 
extrema of any of the parameters that characterize the density
profile of the progenitor core. (We note that the binding energy
at bounce instead of the value at the onset of collapse, 
though different in detail, does not reveal any better correlation.)
While many of the BH formation cases are located in a local
peak of $M_\mathrm{Fe}$, $E_\mathrm{b}$, $M_{\mathrm{O}_<}$, and
$\xi_{2.5}$ between $\sim$22\,$M_\odot$ and $\sim$26\,$M_\odot$,
there are exceptions in this interval. 
Moreover, there are BH formation cases
below and above this ZAMS mass window with clusterings around
15--16\,$M_\odot$, $\sim$18\,$M_\odot$, and 21--22\,$M_\odot$.
Also the C+O-core and O-shell masses, which exhibit a nearly
monotonic increase up to a progenitor ZAMS mass of 28\,$M_\odot$ 
and scattering around a high value beyond, do not show
any correlations with BH formation cases.

The clearest indication of possible BH formation seem to be high
values of $\xi_{2.5}$, at least local maxima (as visible between
15 and 16\,$M_\odot$ and around 18\,$M_\odot$). But there are 
exceptions like progenitors near 21\,$M_\odot$, which have
relatively high values of $\xi_{2.5}$ and $E_\mathrm{b}$.
Despite this fact they explode, correlated with possessing
lower values of 
$M_{\mathrm{O}_<}$ than BH formation cases in the close 
neighborhood. We therefore conclude that the dynamics leading
to a successful SN blast does not depend on a single property
of the progenitor star but on different ones or even a combination
of them.

Although of all quantities displayed in 
Fig.~\ref{fig:progenitorprops} the compactness parameter
$\xi_{2.5}$ in the form of local maxima exhibits the tightest 
correlation with BH formation cases, the corresponding threshold 
value considered by \citet{OConnor2011}, i.e., $\xi_{2.5}>0.45$,
significantly underestimates the BH formation probability 
compared to our results. The progenitor variability seen
in Fig.~\ref{fig:progenitorprops}, however, defies the 
specification of a better and reliable number for the BH 
formation limit. Instead, there is a wide range of $\xi_{2.5}$
values between $\sim$0.15 on the one side and nearly 0.35 on the
other side, for which NS and BH formation cases occur.
Of course, any such conclusions can be affected by the
1D nature of our simulations. It cannot be excluded that
multi-dimensional effects will shift the $\xi_{2.5}$ limit for
BH formation to larger values.

\begin{figure*}
\plotone{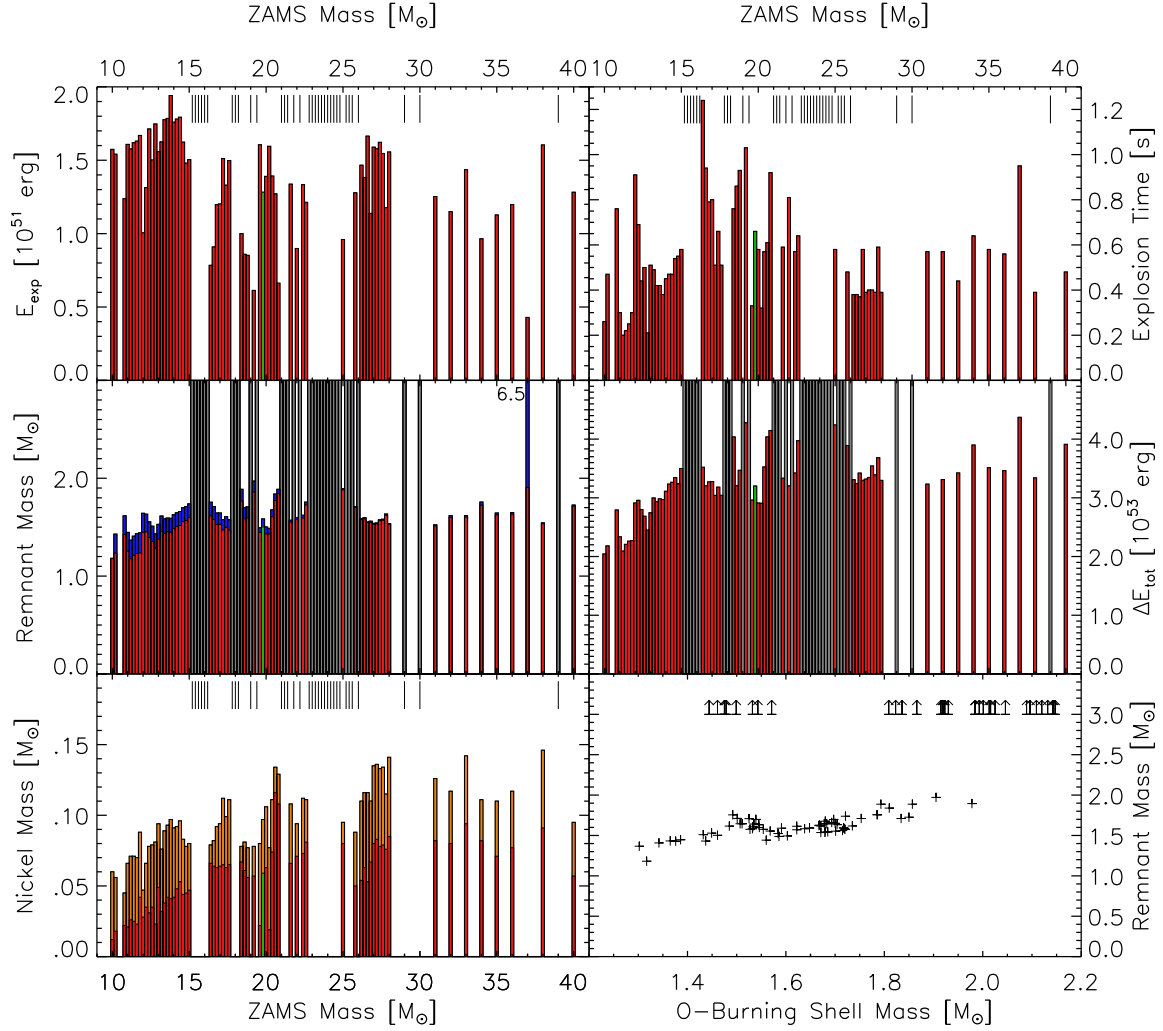}
\caption{Explosion and remnant properties resulting from
our parametrized 1D neutrino-driven SN simulations: explosion
energy ({\em top left}), time of the onset of the explosion
({\em top right}), baryonic  mass of the compact remnant
({\em middle left}), total release of gravitational binding
energy by the compact remnant in neutrinos ({\em middle right}),
and ejected $^{56}$Ni mass ({\em bottom left}) as functions of
stellar birth (ZAMS) mass. The lower right panel shows the
compact remnant mass vs.\ the enclosed mass at the base of
the oxygen-burning shell of the progenitor, where
the stars possess an entropy jump of varying size.
The green histogram
bar indicates the 19.8\,$M_\odot$ calibration model (see text).
While vertical ticks in some panels mark masses where computed
models did not explode, grey histogram bars reaching to the upper
panel edge and arrows in the bottom right panel signal the
formation of a BH containing the whole mass of the progenitor
at collapse. The only exception here is the 37\,$M_\odot$ star,
where the explosion expulses $\sim$3.2\,$M_\odot$
while 4.5\,$M_\odot$ of fallback give birth to
a BH with 6.5\,$M_\odot$. Blue histogram segments indicate
fallback masses and orange segments the uncertainties of the
$^{56}$Ni ejecta masses. The latter uncertainties are associated
with inaccuracies in the $Y_e$ determination of the neutrino-heated
ejecta because of our approximative treatment of neutrino transport.}
\label{fig:SNremnants}
\end{figure*}

\begin{figure}
\plotone{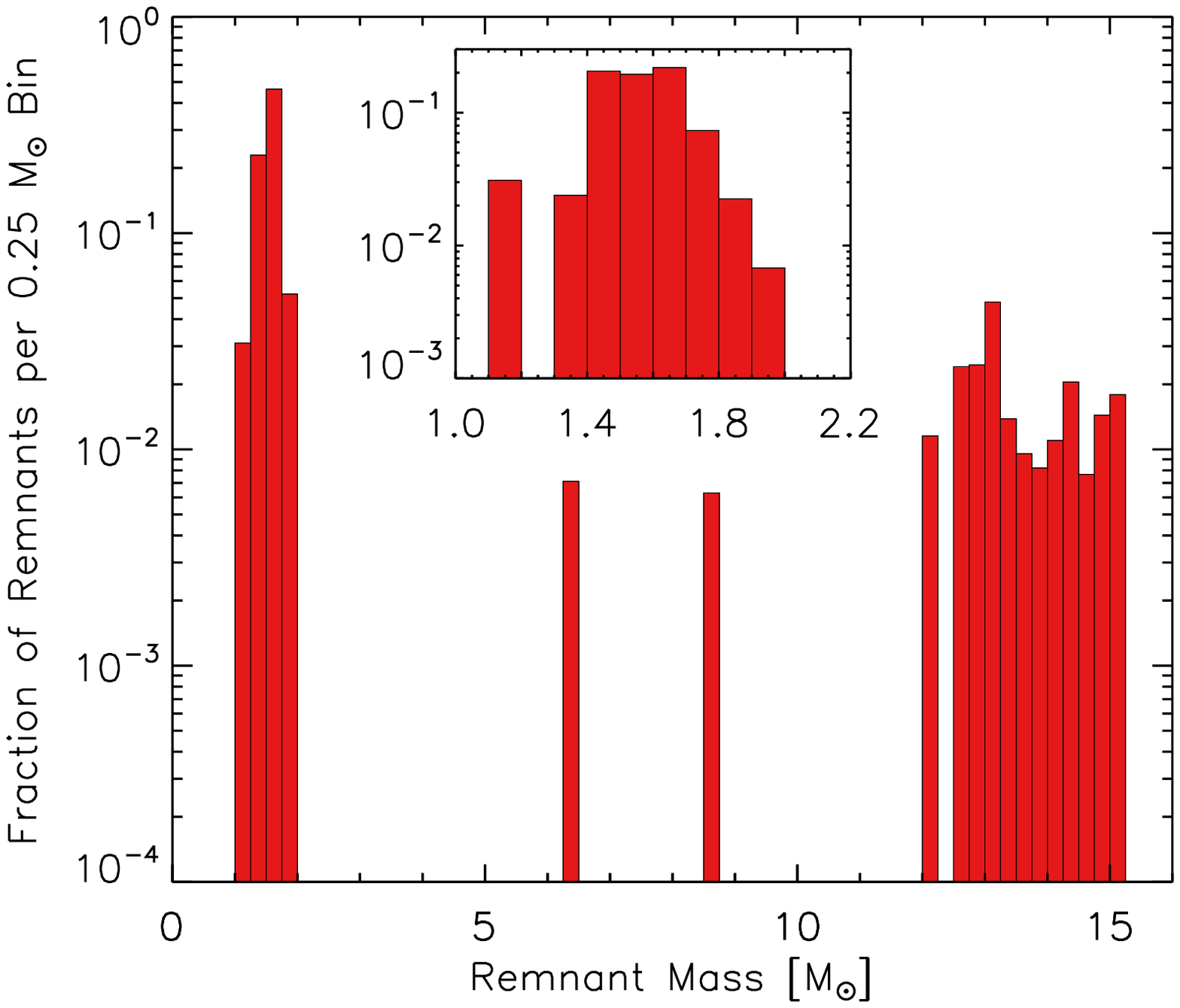}
\caption{Normalized distribution of (baryonic) remnant masses originating
from the neutrino-driven explosion models of our investigated set
of solar-metallicity progenitors, weighted by Salpeter's (1955)
stellar birth rate function. While the main panel shows the fraction
of remnants per 0.25\,$M_\odot$ bin, the insert provides an
enlarged and refined display of the NS mass range with histogram
bars giving the remnant fractions per 0.10\,$M_\odot$ bin.}
\label{fig:remnantmassdistribution}
\end{figure}

\section{Results for explosions and remnants}

The stellar core collapse until $\sim$10--15\,ms after bounce is
computed with the {\sc Prometheus-Vertex} code including sophisticated
neutrino transport \citep{Rampp2002,Buras2006a}. 
At that time the bounce shock has converted to
a slowly expanding accretion shock. We then continue the simulations 
of the subsequent accretion phase and the possible explosion with the 
{\sc Prometheus} version described in Sect.~\ref{sec:numerics}. The
mapping, excision of the NS core, and approximate neutrino treatment
does not cause any worrysome transients.

\subsection{Explosion properties}
 
Explosions can develop in the case of a favorable interplay
of mass accretion rate and neutrino luminosities \citep[e.g.,][]
{Burrows1993,Janka2001,Fernandez2012}. In all 
successful cases compared to failed explosions of neighboring
progenitors, the mass accretion rate is either lower during a long 
postbounce period or decreases rapidly when a 
composition-shell interface arrives at the shock. Shock revival
occurs when the neutrino luminosity is still sufficiently
high (and thus neutrino heating strong enough) at this time.
In a large number of successful and unsuccessful models the
decreasing mass-accretion rate triggers 
shock oscillations, which indicate the proximity
to runaway conditions \citep{Buras2006b,Murphy2008,Fernandez2012} 
and whose amplification also leads to large-amplitude pulses 
of the accretion component of the driving neutrino luminosity
\citep[see][]{Buras2006b}.
In some stars the explosion is fostered by the Si/O
interface reaching the shock relatively soon after bounce,
either due to its location at a smaller mass coordinate or
because of higher mass accretion rates at earlier times,
corresponding to a more compact Si-layer. In this case
the high accretion luminosity seems to be supportive. (More
information on the time evolution, dynamics, and the neutrino 
emission of
our models will be provided in a separate paper). In summary,
the destiny of a collapsing star does not hinge on a single
parameter but depends on the overall structure of the stellar
core.

Figure~\ref{fig:SNremnants} gives an overview of the 
results of our whole model set. All displayed quantities exhibit
considerable scatter even in narrow mass windows, which is a 
consequence of the nonmonotonicities of the progenitor structure.
Failed explosions with BH formation seem to be possible for progenitors
below 20\,$M_\odot$, and successful SNe with NS formation are
found also between 20 and 40\,$M_\odot$. While below 15\,$M_\odot$
all core collapses produce NSs, the investigated progenitor set
yields several ``islands'' with preferred BH creation above 
15\,$M_\odot$. A discussion how BH formation
cases correlate or do not correlate with the density 
structure and characteristic quantities of the progenitor cores
can be found in Sect.~\ref{sec:progenitors}.

The energies of the neutrino-driven explosions do not exceed
$2\times 10^{51}$\,erg and
$^{56}$Ni production up to 0.1--0.15\,$M_\odot$ can be expected.
Note that our determination of nickel yields is uncertain because
the neutrino transport approximation does not provide very accurate
information of the $Y_e$ in the neutrino-heated ejecta and in the
neutrino-driven wind, where a sizable contribution to the ejected
nickel may come from (orange histogram bars in the bottom left
panel of Fig.~\ref{fig:SNremnants} denote the corresponding
production masses of the neutron-rich tracer nucleus in the network).
The onset of the SN blast (roughly defined by the moment the shock 
passes 500\,km) varies in time between $\sim$0.2--1.2\,s after
bounce, thus including ``early'' as well as ``late'' explosions.
Later explosions tend to be less energetic, because less mass is
available in the more dilute gain layer for absorbing energy from
neutrinos. This simple relation for each progenitor is partly
masked by the large differences of the mass infall rates in the
different stars: A dense stellar core allows for a high mass 
accretion rate by the shock and a relatively massive gain layer 
even at later postbounce times.
Moreover, the neutrino-driven wind, which consists
of mass shed off the PNS surface by neutrino heating,
adds a contribution to the explosion power that does not depend
on the density structure of the silicon and oxygen layers.

The blast-wave energy of the SN emerges from a combination
of different positive contributions: 
(a) The total (internal plus kinetic plus gravitational)
energy of the neutrino-heated 
postshock matter when it begins expansion away from the gain radius
after shock revival;
(b) the energy released by the recombination of
nucleons and $\alpha$-particles to heavy nuclei in the expanding
postshock material; 
(c) energy from neutrino absorption in matter that is accreted 
through the shock and channeled toward the gain radius to be 
partly ejected again during a phase of simultaneous accretion
and outflow after the onset of the explosion;
(d) energy carried outward by the neutrino-driven wind from the
PNS surface after the accretion has ended;
(e) energy released from nuclear burning in the shock-heated
ejecta.
Two negative effects reduce the energy that finally escapes
with the SN ejecta:
(f) The binding energy of the stellar layers swept up by the
accelerating explosion shock;
(g) energy extracted from ejection by mass fallback to
the compact remnant.
In our explosion models we fully account for points (a), (b),
(d), (f), and (g), while point (c) was estimated to be
potentially relevant in the multi-dimensional situation
\citep{Marek2009}, and effect (e) typically makes only a small
contribution to the SN explosion energy (the burning of 
0.1\,$M_\odot$ of Si and O releases roughly $10^{50}\,$erg).
Also term (a) is usually unimportant, because
simulations show that the matter in the postshock layer
is marginally unbound (i.e., its total energy is near zero)
when shock revival sets in and the gas begins outward expansion 
behind the accelerating blast wave \citep{Scheck2006,Fernandez2012}.
The dominant positive contributions to the explosion energy in our
1D models are therefore given by terms (b) and (d). The power
of the neutrino-driven wind, point (d), can account
for $\sim$30--70\% of the SN explosion energy \citep[with higher 
relative importance in more energetic explosions;
see the detailed analysis and especially Fig.~C.5 
in][]{Scheck2006}\footnote{Note that 
\citet{Scheck2006} provide
an analysis of the energy contributions only for the first 
second of the explosion. Their results, however, have more
general validity because the positive energy injected to the 
explosion after the first second is in most cases roughly
compensated by the negative energy of point (f).}.

\subsection{Remnant properties}

The NS baryonic masses are in the range of 
$\sim$1.2--2\,$M_\odot$. For the observable gravitational 
masses these numbers need to be reduced by the mass defect
associated with the total neutrino-energy loss during NS formation,
corresponding to $\Delta E_\mathrm{tot}/c^2$. Our model estimates
of $\Delta E_\mathrm{tot}$ are given in the middle right panel
of Fig.~\ref{fig:SNremnants}, implying corrections roughly between
0.11\,$M_\odot$ and 0.23\,$M_\odot$ for the least massive up to the
most massive NSs. However, these numbers are
approximative ---though in a reasonable range---, and exact 
binding energies depend on the NS EoS and have to be determined 
from detailed general relativistic NS structure models. 
The smallest BH contains 6.5\,$M_\odot$ and forms 
in the 37\,$M_\odot$ star by late fallback of 4.5\,$M_\odot$. 
Since the progenitor has a mass of 
$\sim$9.7\,$M_\odot$ at collapse, the SN explosion ejects roughly
3.2\,$M_\odot$. All other BHs
originate from failed explosions and absorb the total
mass of the progenitor at collapse ($>$8.5\,$M_\odot$).
Although the remnant mass follows a trend of growth with the
enclosed mass at the bottom of the oxygen-burning shell
(Fig.~\ref{fig:SNremnants}),
this location is no reliable indicator for the fate of the
star because some models with relatively small silicon core do
not explode. However, BH formation clusters in regions where
local maxima of the NS mass occur.

Note that we do not terminate the neutrino emission from the 
excised core when the mass of the compact remnant exceeds the 
BH formation limit. Instead, the accreting central object is 
considered to remain gravitationally stable, for which reason 
its total energy loss can reach 
$\Delta E_\mathrm{tot} > 5\times 10^{53}$\,erg
(middle right panel of Fig.~\ref{fig:SNremnants}) 
and its cooling time can become 
$t_{90} \gtrsim 6$\,s (Fig.~\ref{fig:pnscoolingtime}).
However, none of our successful explosions was triggered by this 
unphysical, long-time release of neutrino energy, because all
NSs have a baryonic mass of $\le$2\,$M_\odot$. This also holds
for the compact object that is temporarily present (i.e., until 
massive fallback occurs long after the launch of the explosion)
in the 37\,$M_\odot$ progenitor.
The value of the baryonic mass of $\le$2\,$M_\odot$ is
safely below the gravitational mass limit of $1.97\pm 0.04$\,$M_\odot$
set by the recent precision measurement of \citet{Demorest2010},
according to which objects up to at least this mass do not collapse 
to BHs.

A gap that seems to be present in the observed remnant distribution
between NS and BH masses
\citep[e.g.,][]{Ozel2010,Valentim2011,Farr2011} 
is compatible with our results. This is 
visible in Fig.~\ref{fig:remnantmassdistribution}, which displays
the normalized (baryonic) remnant mass distribution originating
from our neutrino-driven explosion models, weighted by Salpeter's 
\citeyearpar{Salpeter1955} stellar birth rate function (with power-law
index $-2.35$). We obtain a relatively wide
distribution of NS masses (see insert). Note that these are birth
masses resulting from single, nonrotating (or at most slowly 
rotating) stars, for which effects like mass transfer and mass loss
associated with possible binary interaction during the progenitor
evolution have been ignored. Accretion of the NSs in binaries will,
of course, also modify this NS birth-mass distribution.

The fallback mass that adds to the initial NS mass is shown in
blue in the middle left panel of Fig.~\ref{fig:SNremnants}.
Fallback is bigger (up to $\sim$0.2\,$M_\odot$) 
for the lower-mass progenitors where an extended 
hydrogen envelope leads to a significant deceleration of the 
explosion shock and a stronger reverse shock. The distribution
of Fig.~\ref{fig:remnantmassdistribution} implies that 22.5\%
of the solar-metallicity progenitors produce BHs at the end
of their lives. We point out that
in the considered sample of 101 solar-metallicity progenitors 
only the 37\,$M_\odot$ star produces a fallback SN with 
delayed BH formation. In this case the black hole swallows
a baryon mass of $\sim$6.5\,$M_\odot$ while the explosion 
ejects the remaining $\sim$3.2\,$M_\odot$ of the pre-collapse
mass of the star. 

Compared to the neighboring, exploding stars of
36\,$M_\odot$ and 38\,$M_\odot$, the 37\,$M_\odot$ progenitor 
possesses a higher enclosed mass $m(r)$ and higher gravitational
binding energy $E_\mathrm{b}(r)$ in the central 15,000\,km.
This is reflected by its larger values of 
$E_\mathrm{b}(M_\mathrm{Fe})$, $M_{\mathrm{O}_<}$, and
$\xi_{2.5}$ in Fig.~\ref{fig:progenitorprops}, although 
$M_\mathrm{Fe}$ is slightly smaller.
For this reason the 37\,$M_\odot$ model explodes only very late
($t_\mathrm{exp}\sim 1$\,s) and with low explosion energy
(Fig.~\ref{fig:SNremnants}). 
Because of the late onset of the blast-wave
acceleration the neutrino energy injected to the explosion
and the shock velocity become considerably lower than in the 36
and 38\,$M_\odot$ cases, in fact too low to unbind the major 
part of the star.
It is interesting to note that the 39\,$M_\odot$
progenitor is even more extreme in all of its characteristic
parameters displayed in Fig.~\ref{fig:progenitorprops}.
Correspondingly, it does not achieve to
explode and collapses to a BH as a whole. In contrast,
the 34\,$M_\odot$ model is also locally extreme in some
of its structural parameters, but its compactness $\xi_{2.5}$
is lower and it explodes much earlier than the 37\,$M_\odot$ 
star. The latter therefore defines an intermediate, seemingly
rare situation in which the combination of a high binding energy
and core compactness of the star and a late onset of the explosion
and low explosion energy favors the fallback of a large fraction 
of the progenitor mass.

\section{Conclusions}

We performed simulations of neutrino-powered explosions 
in spherical symmetry for a large set of solar-metallicity, 
nonrotating stars with Fe-cores. We calibrated an analytic
neutrino-cooling model of the PNS core such that the observed
explosion properties of SN~1987A were reproduced for progenitors
around 20\,$M_\odot$. This yields numbers for the total energy
release and cooling timescale of the newly formed NS in rough
agreement with the SN~1987A neutrino measurement, thus
confirming the overall consistency of our approach.

The neutrino-driven explosions turned out to be fostered by
``steps'' in the stellar density and entropy profiles, which
reduce the mass-infall rate and ram pressure of the accretion
flow and thus allow the shock to expand. Explosions set in 
either by a continuous acceleration of the shock or after a
phase of shock oscillations with growing amplitude. The onset
times range from only 0.1\,s to more than 1\,s after core bounce,
meaning that depending on the progenitor the explosion
can occur relatively early or with a considerable delay.
This conflicts with the claim that neutrino-driven explosions 
should generally be launched soon after bounce in order to
account for the observed gap between highest NS masses and 
smallest BH masses \citep{Belczynski2011}.

Reflecting the ``erratic'' structure differences of the
progenitor models, the explosion properties and remnant
masses exhibit highly nonmonotonic variability even in narrow
ranges of the ZAMS mass. This might explain the considerable
differences of the explosion energies and nickel masses inferred
for observed SNe from lightcurve and spectral analyses as
function of the ZAMS mass of the progenitors \citep[e.g.,][]
{Utrobin2011,Smartt2009a,Smartt2009b}
without invoking additional pre-collapse degrees of 
freedom like rapid core rotation. 
However, the large variations of explosion and remnant properties
even for small changes of the progenitor ZAMS mass
should be taken with caution. The mixing length or mixing length
plus overshooting description of convection used in the employed
progenitor models is extremely unstable to numerical effects in
the late burning stages. While the trends we found in dependence
of iron-core mass, core-binding energy, and compactness are 
likely to be valid, the large structural differences of progenitor
models with similar ZAMS masses probably do not reflect 
physical reality. Instead, late stage convection is subject to
substantial dynamic instability, for which reason there may be
stochastic variations in the core properties at any given mass.
An exploration of this possibility, however, has to be deferred
until multi-dimensional progenitor models, self-consistently
evolved to the onset of core collapse, become available.

Neutrino heating seems
hardly able to provide explosion energies significantly 
higher than $2\times 10^{51}$\,erg and $^{56}$Ni yields
in excess of 0.1--0.2\,$M_\odot$. Observed SNe with
higher explosion energies and nickel production \citep[e.g.,][]{Tanaka2009}
are therefore likely to be driven by a mechanism that is different
from neutrino heating and could involve magnetohydrodynamic processes.

The distribution of NS birth masses (including fallback)
spans from $\sim$1.2\,$M_\odot$ to $\sim$2.0\,$M_\odot$ in 
baryonic mass with a broad maximum between 1.4\,$M_\odot$ and
1.7\,$M_\odot$. BH formation occurs in certain mass intervals
and seems to be possible not only for progenitors with ZAMS
masses beyond 20\,$M_\odot$ but also between 15\,$M_\odot$ and 
20\,$M_\odot$. With a single exception (the 37\,$M_\odot$
progenitor) all BHs swallow the progenitor completely.
This suggests that fallback SNe could be rare cases in 
solar-metallicity environments although roughly 23\% of
the progenitor population end as BHs. Because of the large
mass loss of stars with $M_\mathrm{ZAMS} > 15$\,$M_\odot$,
the fallback is strongest (up to $\sim$0.2\,$M_\odot$) 
for stars below $\sim$20\,$M_\odot$. Correspondingly, we find
a wide gap between the maximum NS birth mass (around 
1.8\,$M_\odot$ gravitational mass) and the minimum BH
mass (roughly 6\,$M_\odot$), compatible with conclusions 
from observed compact remnant masses \citep[e.g.,][]{Ozel2010,Valentim2011,Farr2011}.
It should, however, be noted that our predictions do
not account for binary effects prior and after the 
SN explosion, nor do we include NSs formed from 
ONeMg-core progenitors. A direct comparison with 
observations therefore requires caution.

The compactness parameter $\xi_{2.5}$ introduced by
\citet{OConnor2011} provides the best indicative 
quantity based on the pre-collapse structure for the 
final fate of a star, i.e., whether it is 
likely to end as NS or BH. However, 
our models fail to explode significantly below the
discriminating value of $\xi_{2.5} = 0.45$ considered by
\citet{OConnor2011}. If this result were valid more 
generally, in particular
also for lower-metallicity progenitors, it will
have implications for the discussion of GRB progenitors
\citep{OConnor2011,Dessart2012}.
Instead of depending on an exact bifurcation value of
$\xi_{2.5}$ we see BH formation occurring preferentially
(but not exclusively) in regions with local maxima of
$\xi_{2.5}$ and NS as well as BH formation to be possible
for $\xi_{2.5}$ between $\sim$0.15 and $\sim$0.35. 
The success or failure of 
the explosion mechanism obviously depends on the 
progenitor structure (determining the time-dependent
mass-accretion rate, 
shock radius, and accretion luminosity) in a complex way 
and cannot be predicted exactly on grounds of a single 
parameter.
 
So far we have investigated only solar metallicity
progenitors from one modeling group, but we plan to 
extend our studies also on sets of progenitors from other
groups and for different metallicities. Naturally,
the spherical symmetry of our simulations is a 
major constraint, since the explosion mechanism has
been recognized to be intrinsically multi-dimensional
\citep[e.g.,][]{Janka2007,Burrows2007}.
Nonradial fluid flows in the gain layer have a supportive
influence and thus reduce the critical luminosity for 
driving the explosion \citep[e.g.,][]{Janka1996,Murphy2008,
Nordhaus2010,Hanke2011,Murphy2011}. 
With a higher neutrino-heating
efficiency lower values of the early postbounce luminosities 
from our analytic PNS core-cooling model would be needed to
reproduce the explosion properties of SN~1987A. This, in turn,
would stretch the PNS cooling timescale, which in our 
simulations tends to be on the lower side of the range
compatible with the SN~1987A neutrino data.

Nevertheless, this would only mean a recalibration of 
the time decay of the core-neutrino source adopted
for our modeling, but it is unlikely to fundamentally 
alter qualitative aspects of the behavior of the different
progenitors in relative comparison to each other. 
Therefore we are hopeful that our basic 
conclusions will retain their validity also in the 
multi-dimensional case even if quantitative changes
(for example lower explosion energies by less powerful
neutrino-driven winds in low-mass progenitors) could be
the consequence. It also cannot be excluded that some of
the extreme sensitivity of 1D explosions to details of the
progenitor structure might be moderated or disappear
in multi-dimensional models. It is interesting to note,
however, that the relatively early explosions of the 
11.2\,$M_\odot$ and 27\,$M_\odot$ progenitors of our model
set and the later explosion of progenitors around 15\,$M_\odot$ 
are fully compatible with recent results of 2D
models with sophisticated neutrino transport 
\citep[cf.][]{Marek2009,Muller2012a,Muller2012b}.

Our findings challenge canonical thinking, e.g.\ of the progenitor
masses leading to BH formation \citep[e.g.,][]{Zhang2008,OConnor2011},
and they raise doubts about claims in the 
literature that the possible gap between NS and BH masses 
requires explosions that set in relatively early after core bounce
\citep{Belczynski2011,Fryer2012}. We have shown
that a highly variable onset time of the explosion and little
fallback for massive progenitors (i.e., a low occurrence 
rate of fallback SNe) naturally leads to the mass gap.
On the other hand, a long delay of the beginning of the 
explosion is questioned by the recent Bayesian analysis of
the NS mass distribution in double NS systems by \citet{Pejcha2012}.
These authors conclude that for solar metallicity stars the 
explosion most likely develops near the edge of the iron core
at a specific entropy of $S/(N_\mathrm{A}k_\mathrm{B}) \approx 2.8$.
It is hard to understand how the explosion mechanism could
dynamically couple to this specific pre-collapse value of the entropy,
because the core density and entropy profiles are highly variable
for the different progenitors, they are modified by the nuclear
burning during the collapse phase, and the neutrino-driven
mechanism in spherical symmetry as well as in the multi-dimensional
case is sensitive to the mass-infall rate rather than the pre-collapse
entropy structure. For these reasons the mass-accretion rate and
the accretion luminosity exhibit large differences between different
progenitors when compared for specific values of the entropy, for
which reason a wide range of explosion times appears plausible.
Interestingly, however, \citet{Pejcha2012} favor a NS formation 
scenario with no or very little mass fallback, which is consistent
with the tendency of our models compared to piston-driven 
explosion models. Correspondingly, the probability distribution
computed by \citet{Pejcha2012} with
our predicted remnant masses has little power for NS masses above
$\sim$1.6\,$M_\odot$ even when our fallback masses are included.
Nevertheless, the theoretical distribution is shifted by 
$\sim$0.1\,$M_\odot$ to higher masses relative to the observational
double NS data. If this mismatch cannot be attributed to an
overestimated delay time of the explosion in our supernova models
as argued above, it might be connected to the low-number
statistics of the empirical data or might point
to smaller iron-cores than present in the solar-metallicity
progenitors of our study.

Certainly, our simulations can mean only a very first step of
systematically exploring the consequences of neutrino-powered
explosions for a wide variety of progenitors. Nevertheless,
our results already shed new light on a number of important 
astrophysical questions connected to how the properties of SN 
explosions and their remnants connect to those of the progenitor
stars. While the fundamental dependences of explosion 
and remnant properties on the progenitor structure are likely to 
possess more general validity, the detailed functional variation 
with the progenitor mass is subject to a number of caveats.
These caveats are, in particular, the employed set of progenitor 
models with their large structural
variations even for similar ZAMS masses, the calibration of
the neutrino-cooling models of the PNS core based on a 
certain choice of the SN~1987A progenitor model, and the 
constraint of the explosion modeling to spherical symmetry.
Future work will have to explore the consequences of these
constraints and deficiencies.

\acknowledgements
We thank Lorenz H\"udepohl and Annop Wongwathanarat for support.
At Garching,
this work was supported by the Deutsche Forschungsgemeinschaft through
Sonderforschungsbereich/Transregio~27 ``Neutrinos and Beyond'',
Sonderforschungsbereich/Transregio~7 ``Gravitational-Wave Astronomy'',
and the Cluster of Excellence EXC~153 ``Origin and Structure of the
Universe''. A.A.\ acknowledges funding by the Helmholtz-University
Young Investigator Grant No. VH-NG-825. The computations were partially
performed at the Rechenzentrum Garching.

\bibliographystyle{apj}

\end{document}